\documentclass{elsart}
\usepackage[dvips]{graphicx}
\usepackage{amssymb}

\begin{document}
\begin{frontmatter}

\title{
STAR Barrel Electromagnetic Calorimeter Absolute Calibration Using
``Minimum Ionizing Particles'' from Collisions at RHIC
}

\author[WSU]{T.~M.~Cormier}, 
\author[WSU]{A.~I.~Pavlinov}, 
\author[WSU]{M.~V.~Rykov}, 
\author[WSU]{V.~L.~Rykov\thanksref{CA}}, 
\author[IHE]{K.~E.~Shestermanov}
\address[WSU]{Wayne State University, Detroit, MI 48201, USA}
\address[IHE]{IHEP, Protvino, Moscow District 142284, Russia}
\thanks[CA]{Corresponding author. Phone: 
(313)--577--2781; fax: (313)--577--0711;
e--mail: rykov@physics.wayne.edu}

\begin{abstract}
   The procedure for the STAR Barrel Electromagnetic Calorimeter
(BEMC) absolute calibrations, using penetrating charged particle hits
(MIP-hits) from physics events at RHIC, is presented. Its systematic
and statistical errors are evaluated. It is shown that, using this
technique, the equalization and transfer of the absolute scale from
the test beam can be done to a percent level accuracy in a reasonable
amount of time for the entire STAR BEMC. MIP-hits would also be an
effective tool for continuously monitoring the variations of the BEMC
tower's gains, virtually without interference to STAR's main physics
program. The method does not rely on simulations for anything other
than geometric and some other small corrections, and also for
estimations of the systematic errors. It directly transfers measured
test beam responses to operations at RHIC.
\end{abstract}

\end{frontmatter}

\section*{Introduction}
\label{sec:intro}

   The calibration of calorimeters in collider experiments, using
external beams, is typically either impractical or even impossible. As
a result, the calibration is usually performed {\em in situ}, using
several independent complimentary methods to make sure the calibration 
constants obtained by different methods are mutually consistent within
the systematic and statistical uncertainties of each methods. In most
large collider experiments, it is quite common to use radioactive
sources, cosmic rays, and calorimeter hits from high statistics
physics events with clear signatures and hit
patterns~\cite{aleph:90,zeus:1}. For electromagnetic calorimeters
(EMC), examples of the latter include energy-momentum matching for
electrons as they are measured in the tracking detectors and
calorimeter, measuring the positions of invariant mass peaks for known
resonances and short-lived particles ($\pi^{o}$, $\eta$, $J/\psi$,
$Z^{o}$,~\ldots), etc. Most of these processes can and will eventually
be used at various stages of the running STAR experiment for precise
calibrations of its electromagnetic calorimeters in various energy
intervals. However, the ``direct'' calibrations over a wide energy
range, using the physical processes above with electrons and/or
photons in final states, will require a rather long time to obtain
sufficient statistics and to do the appropriate data
analysis. Therefore several ``indirect'' complimentary methods, which
are described in Ref.~\cite{emcfdr:98}, will be used first to set the
channels' gains to the few percent level and obtain the initial
calibration constants immediately after a module installation during
the few first days or even hours of running RHIC.

   In this paper, we will focus on just one of many indirect
approaches which is currently considered as the base-line
method~\cite{emcfdr:98} for the STAR Barrel Electromagnetic
Calorimiter (BEMC) calibration, equalization and continuous gain
monitoring. This method relies on measuring the BEMC towers' responses
to penetrating charged particle hits from physics events at the
running RHIC and comparing them to the ones obtained in the BEMC
towers exposure to an external test beam.

\section{STAR Barrel Electromagnetic Calorimeter}
\label{sec:bemc}

The STAR BEMC, a sampling scintillator-lead calorimeter with the
pseudo-rapidity coverage of $-1 < \eta < +1$\, and $2\pi$\, in azimuth
$\varphi$, has been described
elsewhere~\cite{emcfdr:98,star:94,llope:96}. It will be
installed within the STAR magnet at the radius of $\sim$220~cm just
outside the Time Projection Chamber (TPC) and Central Trigger Barrel
(CTB).
\begin{figure}[htb]
\parbox{4.5cm}{
\centerline{\hbox{
\includegraphics[width=4.5cm,bb=5 240 580 600]{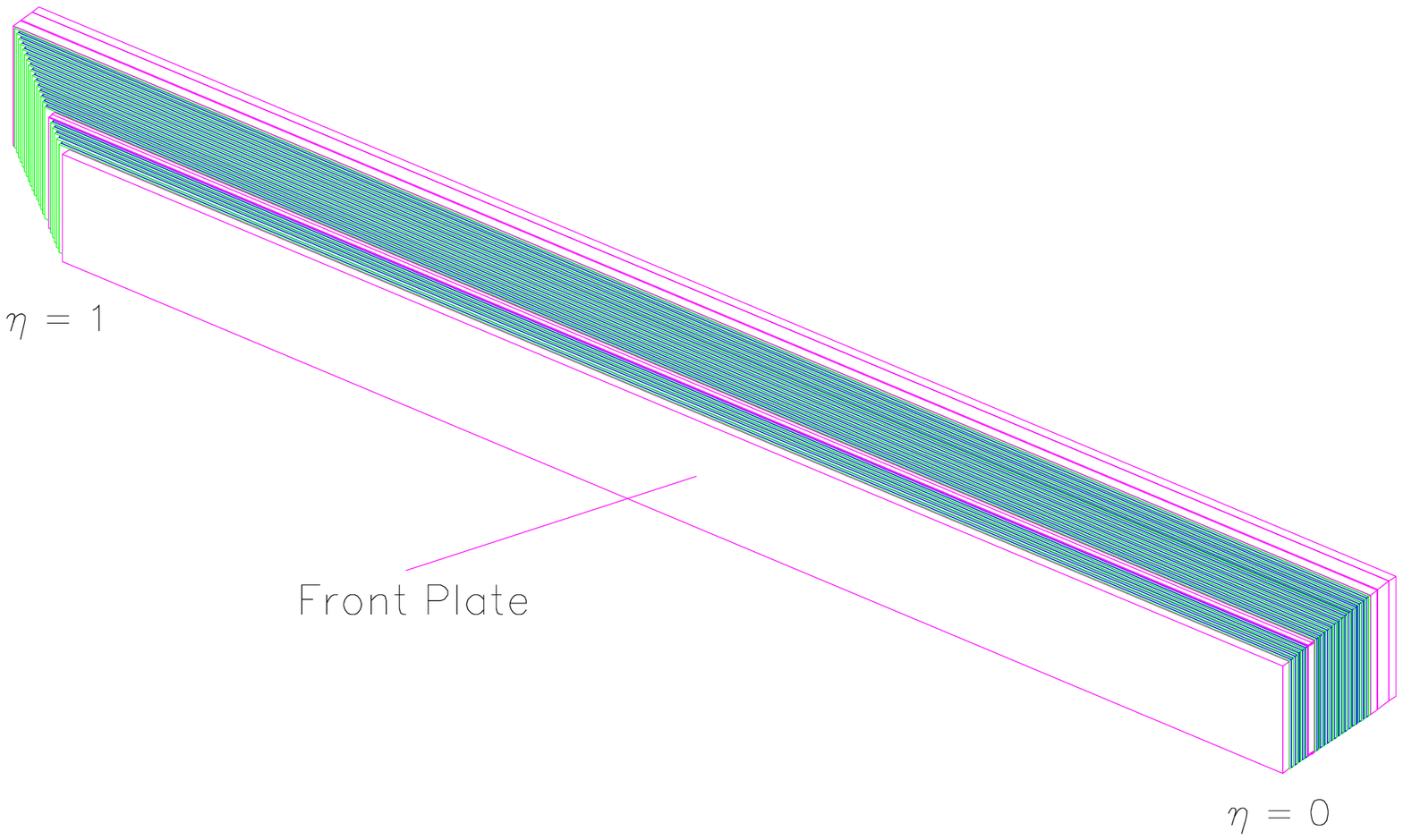}
}}
} \
\parbox{4.5cm}{
\centerline{\hbox{
\includegraphics[width=4.5cm,bb=5 145 560 670]{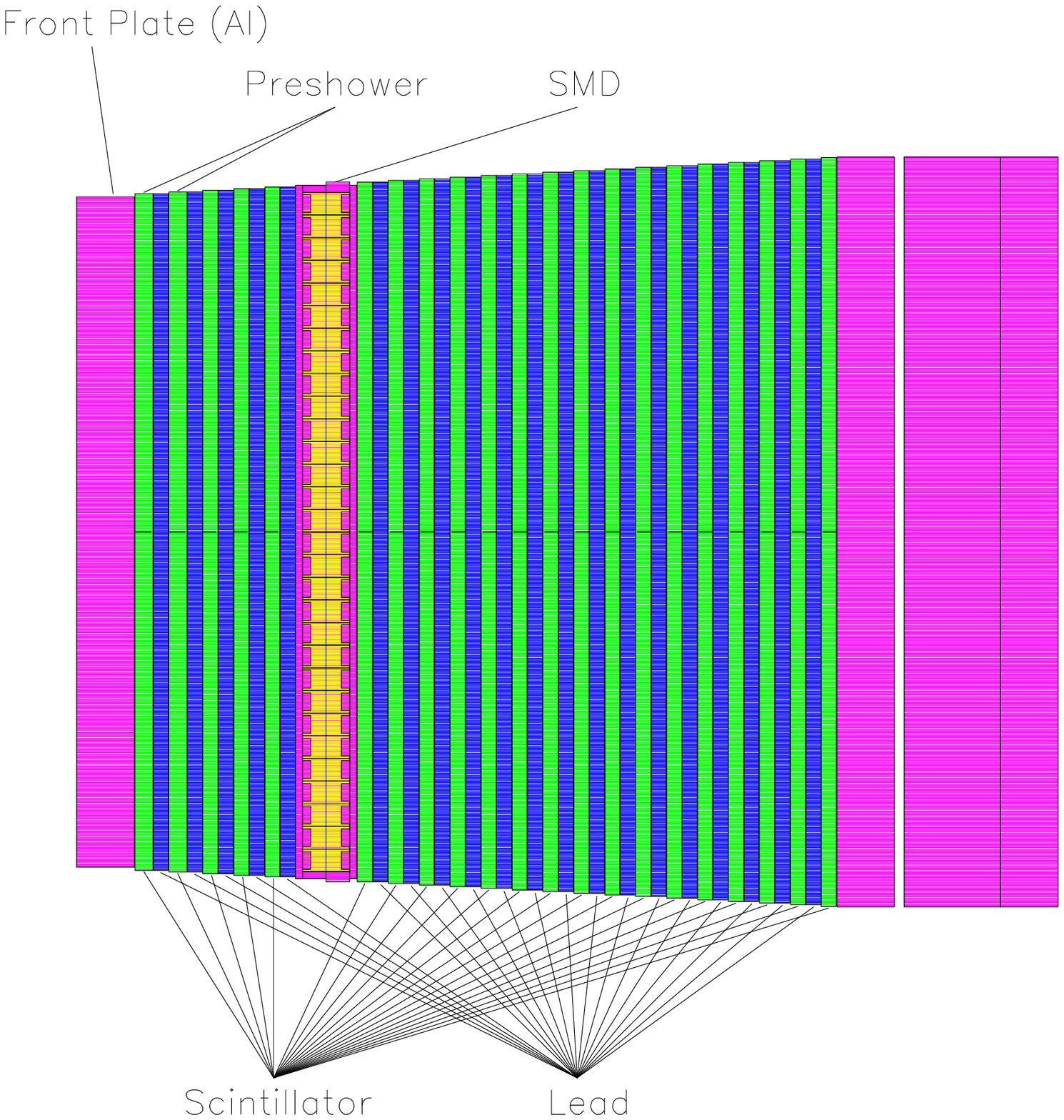}
}}
} \
\parbox{4.5cm}{
\centerline{\hbox{
\includegraphics[width=4.5cm,bb=83 133 530 659]{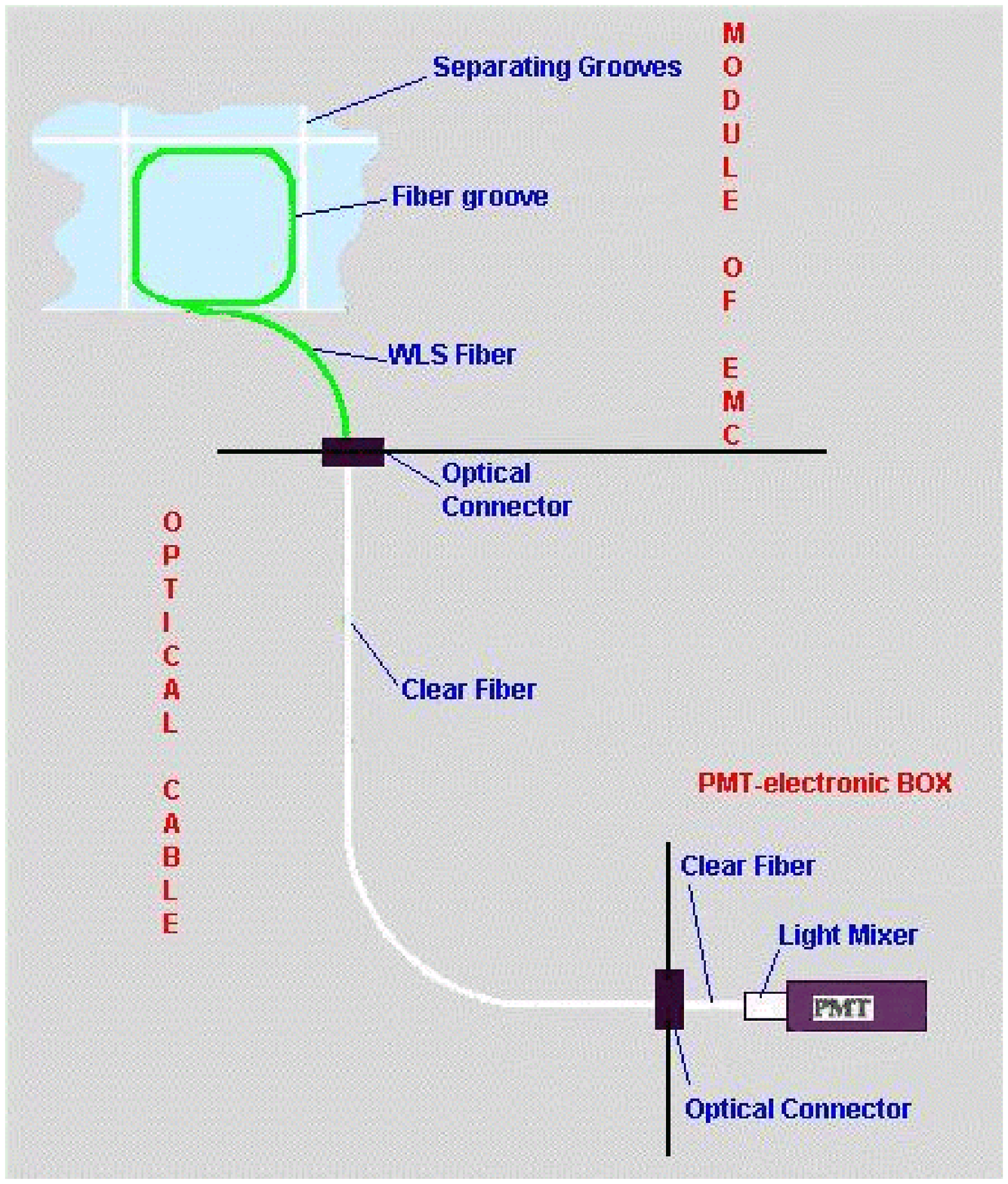}
}}
}
\caption[]
{
\footnotesize
STAR BEMC module view (left), its cross section (center), and light
collection scheme(right).
}
\label{fig:emcmod}
\end{figure}

The BEMC consists of 120 modules of the size
$\Delta\eta\times\Delta\varphi\simeq$ 1$\times$0.1 each. The BEMC
module (Fig.~\ref{fig:emcmod}) is a stack of 21 scintillator megatiles,
separated by 5~mm lead sheets. Each megatile is divided into 40
light-insulated tiles, 20 in $\eta$\, by 2 in $\varphi$, with a
separate readout from each tile, using wave-length shifting fibers
connected to $\sim$3~m long clear fibers. The light from each set of
21 tiles of different layers, which creates a single projective towers
of the size $\Delta\eta\times\Delta\varphi\simeq$ 0.05$\times$0.05, is
collected at a single phototube (PMT), one per tower, totaling in
4,800 channels for the full BEMC coverage. The two front tiles of each
tower, with a second set of embedded fibers to transport the light to
a separate photodetector, provide a ``Preshower'' (PRS) signal, which
is used to enhance $e$-$\gamma$/hadron separation in the
BEMC~\cite{lecompte:306}. To compensate $\sim$20\% loss of light due
to a second fiber, the two front megatiles were made 6~mm thick, while
the thickness of the remaining 19 tiles is 5~mm each. For
perpendicular tracks, the total BEMC depth is $\sim$18 radiation
lengths ($\sim$18X$_{0}$) and $\sim$0.8 of the nuclear absorption
length. A position-sensitive two-layer gaseous Shower-Maximum Detector
(SMD) is placed in front of the 6th megatile at the depth of
approximately 5X$_{0}$\, within BEMC stack.

\section{Requirements to the  STAR BEMC calibrations}
\label{sec:requirements}

   The most stringent requirements on the {\em equalization} of
the BEMC towers come from the needs of the lowest-level, fast
L0-trigger. Well equalized towers are those, having equal
responses, in terms of digitized signals, to electron-photon
hits of, for example, equal energy, $E$, or equal transverse energy
\mbox{$E_{T}=E\cdot\sin\theta$}, where the polar angle \mbox{$\theta =
2\times\arctan(e^{-\eta})$}. The usual BEMC contribution to the fast
trigger is to select events with a high local and/or global energy
deposits\footnote{Either $E$, or $E_{T}$, or both.}
above the chosen L0-trigger's threshold(s). Along with other factors,
the ``sharpness'' of thresholds for this kind of
\mbox{``high-$E_{T}$''} triggers depends on how well towers' responses 
have been equalized. On the other hand, due to a finite BEMC energy
resolution, which also contributes to the widening of thresholds, it
does not make sense to equalize tower gains much better than an
intrinsic BEMC energy resolution. In the range of interest for L0's
$E_{T}$-thresholds in STAR from $\sim$3--5 to
10~GeV~\cite{l0trig:94:96}, the expected intrinsic energy resolution
is \mbox{$\sigma_{E}/E \simeq$
5--10\%}~\cite{emcfdr:98,llope:96}. This means that the {\em
equalization} of the BEMC towers to the level of $\sim$3--4\% is
sufficient in the $E_{T}$-interval of $\sim$3--10~GeV.

   It would also be correct to compare acceptable statistical
uncertainties of the BEMC towers' {\em relative calibrations} to an
intrinsic BEMC energy resolution. The relative tower calibration,
which is the knowledge of its calibration coefficient relative to some
base gain is equivalent, to some extent, to the tower's
equalization but for the data analysis rather than L0-trigger, because
the uncertainties in the calibration coefficients will effectively
contribute to the final BEMC energy resolution. The other difference
is that, while the equalization of towers' responses is important in a
rather limited energy interval of L0-trigger thresholds, the
requirements to the relative calibrations are relevant to the entire
$E_{T}$-range of interest for STAR from $\sim$0.5--1~to
$\sim$50--60~GeV. At energies below $\sim$10~GeV, the knowledge of
the calibration coefficients with the same statistical uncertainty of
$\sim$3--4\%, as for equalization of digitized responses, would be
sufficient. Simulations with GEANT~\cite{geantguide} have shown that,
at the energies above 20--25~GeV, the intrinsic BEMC energy resolution
is almost constant and equal to $\sim$3\%. With this resolution, for
example, the width (RMS) of $Z^{0}$-peak, $\Delta M_{Z}/M_{Z}$, in
$e^{+}e^{-}$-decay mode, including its natural width, will be
$\simeq$3.5\%. For two times worse BEMC energy resolution of 6\%, the
width of $Z^{0}$-peak would increase by a factor of $\sim$1.5 to
$\Delta M_{Z}/M_{Z} \simeq$ 5\%. The summary of this consideration is
that the requirement on the accuracy of the {\em relative
calibration} for each single BEMC tower of no worse than 3--4\% seems
to be valid for the entire energy range from \mbox{$\sim$0.5--1} to
\mbox{$\sim$50--60~GeV}, particularly early in the program. Of course,
over time, one can and will do better, utilizing directly
electron-photon hits. But the point is that relative calibration
coefficients with the level of uncertainty above will not
significantly impact the physics program.

   In STAR, the most restrictive requirements for the {\em absolute
BEMC calibration\/} arise from the measurements of differential cross
sections, which fall steeply with transverse momentum, $P_{T}$. Fits
to the SPS data \cite{ua1ua2inc} for inclusive direct-$\gamma$ and
$\pi^{0}$\, productions at \mbox{$P_{T} >$ 10~GeV/c} give the
dependence of \mbox{$d\sigma/dP_{T} \propto P_{T}^{-(5-5.5)}$}. The
ISR measured spectra~\cite{isr:82} at lower $P_{T}$'s from
$\sim$4~GeV/c fall even more sharply, \mbox{$\propto
P_{T}^{-(6-8)}$}. To measure these differential cross sections with
systematic errors of no more than \mbox{$\sim$10--20\%}, the BEMC 
{\em absolute scale} in the region of interest has to be known at the
accuracy of \mbox{$\sim$1.5--2\%}.

   The requirements on {\em monitoring the variations of calibration
coefficients} over time are directly related to the considerations 
above. Tracking variations of a mean gain for the entire BEMC or its
any large patch\footnote{For example, 40 towers of a BEMC module
or 120 towers of an $\eta$-ring.} has to be done at the accuracy of
the absolute scale, i.e. at about \mbox{1.5--2\%}. The statistical
errors for tracking gain variations for each single tower can be
larger, \mbox{$\sim$3--4\%} as for the relative calibration and
equalization.

   It is also appropriate to mention here the requirement for an
{\em initial setting of PMT gains} after a BEMC module installation in
STAR. Formally speaking, the method of ``MIP-calibrations'', which is
the main subject of this paper, requires only that initial gains be
set so that MIP peaks\footnote{See Secs.~\ref{sec:mipcalib} and
\ref{sec:mip_peak} for all definitions, related to the MIP-calibration
method.} from penetrating charged particles are sitting somewhere
within digitizers' (ADC) ranges, reasonably far from their lower and
upper limits. Then, only in a few hours of running RHIC, the towers
can be equalized to the accuracy of a few percents. However, so as to
not fully incapacitate during these few hours the functionality of
those triggers, which rely on the BEMC signals, it would be desirable
to have the newly-installed module more or less equalized from the
very beginning. The initial equalization at \mbox{$\sim$10--15\%}
seems both sufficient and practically feasible by, for example,
measuring the towers' responses to cosmic rays just before a new
module installation and then transferring these measurements to STAR,
using the Light-Emitting Diode (LED) technique~\cite{emcfdr:98}.

\section{Calibration scheme}
\label{sec:mipcalib}

   Many charged hadrons (along with small admixtures of electrons and
muons) will be produced in every collision at RHIC. In the central
region covered by the STAR BEMC, these are mostly pions. When striking
the BEMC, a significant fraction, $\sim$30--40\% of high energy
charged hadrons, do not deposit a large amount of energy via nuclear
interactions, instead depositing only \mbox{$\sim$20--30 MeV} of
energy in the calorimeter's 21 scintillator layers due largely to
electromagnetic ionization. In this paper, we will loosely call all
these charged hadrons ``Minimum Ionizing Particles'' (MIP),
producing ``MIP-hits'' in the BEMC towers, resulting in ``MIP peaks''
in the signal spectra. For relativistic particles, the position of
MIP peak is nearly independent of momentum and particle species. This,
along with the high yield of charged hadrons, makes it attractive to
explore the feasibility of using high energy MIPs for the
equalization, calibration, and continuous monitoring variations of the
BEMC towers' gains. 

   The calibration scheme using MIP-hits has two stages. In the first
stage, a sample of BEMC modules is exposed to an external beam, for
example at the Alternating Gradient Synchrotron (AGS) of Brookhaven
National Laboratory. The composition of the AGS B2-Line negative test 
beam is mainly $\pi^{-}$'s with some fractions of other hadrons,
electrons and muons of a chosen momentum, selected in the range from
\mbox{0.3--0.5} to \mbox{7--8~GeV/c}. Therefore the ratios of each
tower responses to MIP's and electron's hits,
$MIP/e$-ratios\footnote{\label{footnote:e}In this definition,
``$MIP$'' is the MIP-peak position which is described more precisely
later in the paper. The notation ``$e$'' is used here for the ratio
$S_{e}(E_{e})/E_{e}$, where $S_{e}$\, is the mean BEMC signal from
electrons of energy $E_{e}$. Thus, the $MIP/e$-ratio represents the
energy of electrons which would generate in the BEMC the same mean
signal as MIPs.}, are measured simultaneously (or almost
simultaneously for electrons and MIPs of different momenta). This
makes these measurements completely independent of the actual PMT's
and ADC's gains, possible attenuation and distortion of signals in
cables, delay lines, etc., i.e. in the equipment which might be
necessary for the test beam run but won't be present in the STAR
detector, and vice versa. In the second stage, after the modules have
been installed in their places in the STAR, and the RHIC accelerator
is producing collisions, samples of MIP-hits of particle composition
and momenta as close as possible to those in the test beam are
accumulated for each tower, and the positions of the resulting MIP
peaks are measured. This step essentially completes the procedure of
transferring beam-test results to STAR. For those towers exposed to
the test beam, their responses to electron hits\footnote{Of the
momenta actually used in the test-beam run.} can immediately be
predicted, using the known $MIP/e$-ratios that have been measured at
the beam-test stage. For all other modules, these ratios are
expected to be close to those of the tested  ones, provided key
design tolerances are kept at the module manufacturing stage.

   It is clear that charged particles of low transverse momenta are
useless in the MIP calibration process because their deflection in the
STAR's magnetic field causes them to enter the BEMC at large
angles. Only a small fraction of these particles, if any, pass through
all 21 scintillator tiles of a single tower. On the other hand, if a
\mbox{$P_{T}$-threshold} is chosen too high, the useful event rate
would be too low because of the steep drop of the particle yield as
$P_{T}$\, increases. Simple estimates suggest MIPs of $P_{T}$'s just
above \mbox{$\sim$1~GeV/c} as a good compromise between yield and
utility. In the STAR magnetic field of 0.5~T, the trajectories of no
fewer than \mbox{$\sim$50--60\%} of charged particles with
\mbox{$P_{T}\geq$~1~GeV/c}, produced at the primary
vertex\footnote{And not interacted strongly in the BEMC.}, will pass
through all 21 tiles of a single BEMC tower. An additional important
benefit is that the momenta of \mbox{$\sim$1--2 GeV/c} are within the
range of the AGS test beam.

   The calibration errors of the procedures described above depend on
a number of factors. In the rest of this paper, we will discuss in
details the most important of them.

\section{Design tolerances on the light yield variations from the
scintillator tiles within a tower}
\label{sec:design}

   The calibration procedure above relies on the assumption that the
measured $MIP/e$-ratios from a few modules, exposed to the beam in a
limited energy interval, could be extended to the entire energy range
of interest and also applied to the non-beam-tested modules. To extend
the calibrations to a wider energy range, one has to be sure that
towers' responses were linear or, alternatively, the functional
dependences\footnote{Not accounting scaling factors.}
$S_{e}(E_{e})$\, were known. An application of the beam-test
calibration data to non-tested modules requires that the towers of the
same $\eta$\, in all modules have similar functional dependences
$S_{e}(E_{e})$\, to within \mbox{$\sim$3--4\%} if the specification on
relative towers' calibration is to be satisfied. This in turn implies
certain requirements on the uniformity of the module design and
manufacturing\footnote{It is worth mentioning here that unknown
nonlinearities of the electronic PMT-ADC chains, in the beam-test as
well as in STAR, will go directly to the systematic errors of
MIP-calibrations. However, these nonlinearities are relatively easily
measured and accounted in the calibrations.}.
\begin{figure}[htb]
\parbox{7.4cm}{
\centerline{\hbox{
\includegraphics[width=7.3cm,bb=20 160 535 650]{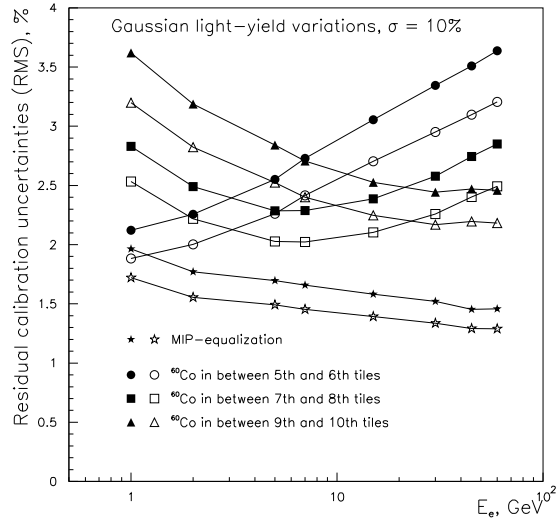}
}}
} \
\parbox{6.0cm}{
\caption[]
{
\footnotesize
Residual uncertainties of calibration coefficients after the tower
equalization, using MIPs and $^{60}Co$\, $\gamma$-source, placed at
various locations in the stack. Solid markers: no tile selection. Open
markers: only tiles with light yields within $\pm$20\% from the mean
value selected at the quality assurance stage.
}
\label{fig:lightyield}
}
\end{figure}

   The functional dependence of a sampling calorimeter response,
in general, and its linearity, in particular, quite strongly depend on
the uniformity of light yields from scintillator layers within a
tower. Individual tower variations of light yield from tile-to-tile
introduce nonlinearities in the energy responses due to the
development of mean shower depth with energy\footnote{See, for
example, Ref.~\cite{tsai:95} for details.}. Therefore, setting limits
during module production on the light yield variations from layer to
layer represents the most critical issue for the maintenance of the
BEMC calibration and equalization within specified limits over a broad
energy range. On the basis of simulations with GEANT, which are
illustrated in Fig.~\ref{fig:lightyield}, it has been determined that,
after towers' equalization with MIPs, the residual uncertainties of
calibration coefficients won't exceed the design limits if light yield
variations were kept within gaussian $\sigma\leq$~10\%. From
Fig.~\ref{fig:lightyield}, it also follows that the equalization,
using MIPs, is expected to provide slightly better results, compared
to another widely used method with $^{60}Co$\,
\mbox{$\gamma$-source~\cite{zeus:1}.}

\section{Reconstructing MIP-peak positions in the beam tests and in
STAR at RHIC}
\label{sec:mip_peak}

   The compositions of MIPs in the beam-test and in STAR at RHIC will
certainly be different and won't be exactly known. To accumulate
sufficient statistics in a reasonably short time, the momentum range
of the selected MIPs in STAR cannot be made as narrow as it was in the
beam test. Moreover, STAR's 0.5~T magnetic field will change
$MIP/e$-ratios compared to those measured at the external beam. All
these differences require introducing multiple corrections and
constitute a potential source of systematic errors. Due to substantial
backgrounds under and in the vicinity of MIP peaks, their positions
are to be determined from some, not so obvious fits to signal
spectra. The actual background depends on the environment and hit
selection. One expects different backgrounds in the module beam test
and in STAR at RHIC. This requires that MIP-peak fits to be, to a
certain degree, robust against background variations, variations of
the total light yields from otherwise identical towers and electronics
noise.

\begin{figure}[hp]
\centerline{\hbox{
\includegraphics[width=13.5cm,bb=20 160 530 660]{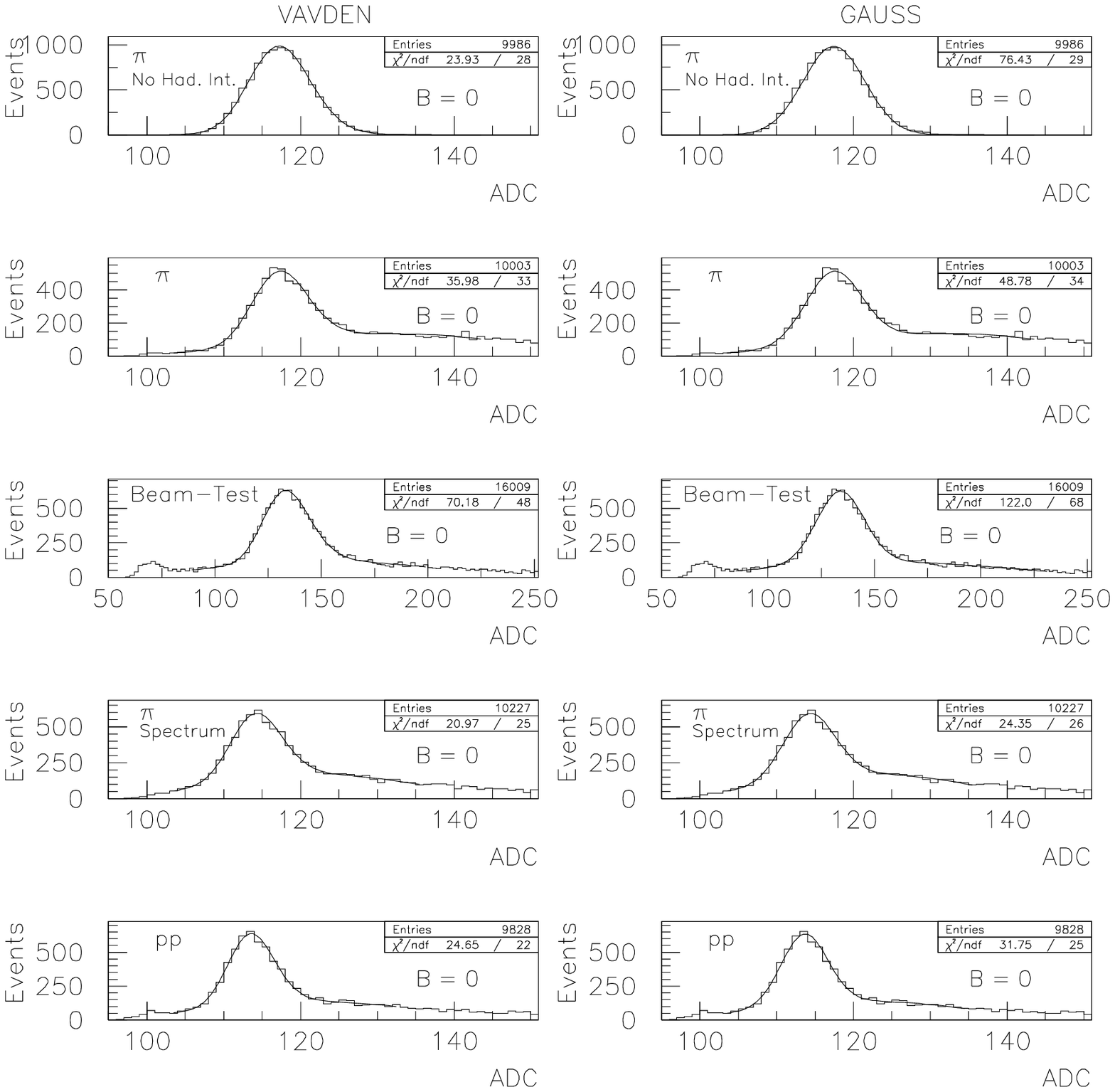}
}}
\centerline{\hbox{
\includegraphics[width=13.5cm,bb=20 460 530 650]{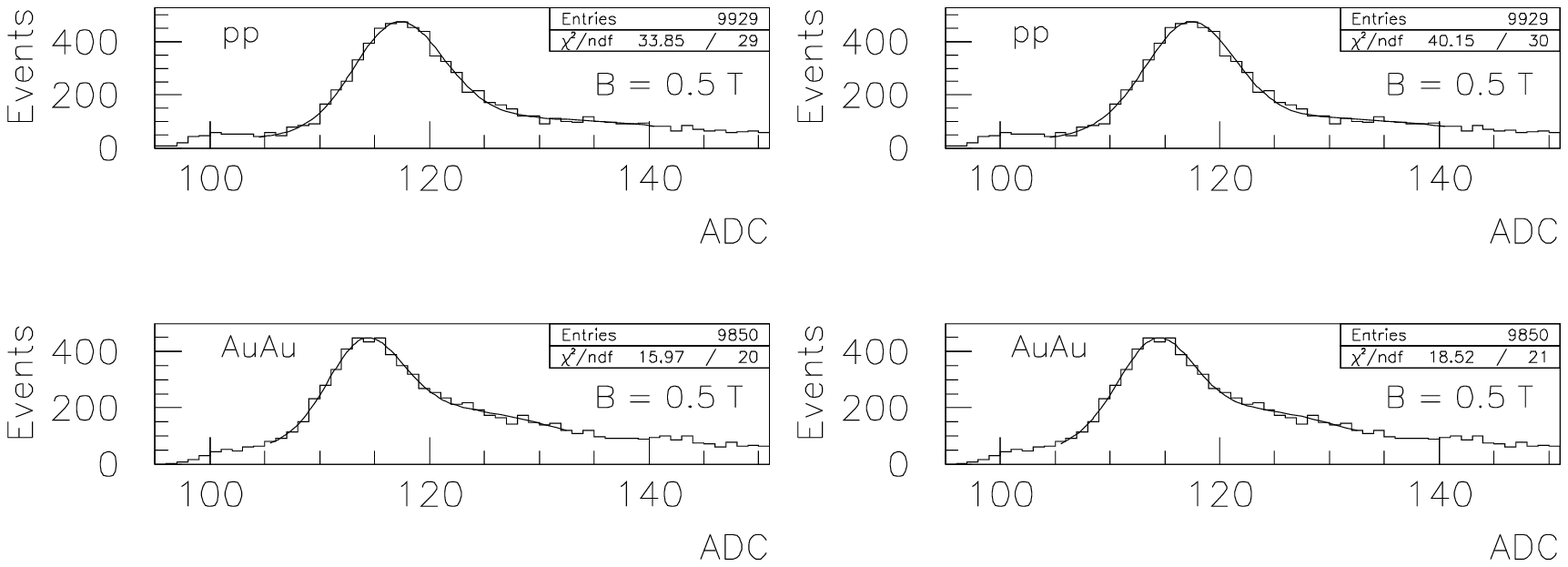}
}}
\caption[]
{
\footnotesize
Fits to the typical PMT signal distributions, using
\mbox{{\em VAVDEN}-based} functions (left column) and gaussians (right
column). Row~1~(top)~--~simulated mono-momentum pions without hadronic
interactions; Row~2~--~simulated mono-momentum pions with hadronic
interactions;
Row~3~--~experimental data from the BEMC Beam-Test-'98~\cite{bt98:01};
Row~4~--~simulated signals from pions with PYTHIA-HIJING's spectrum
\mbox{$dN_{\pi}/dP_{T}\propto\exp[-P_{T}/($0.3 GeV/c$)]$} at
\mbox{$P_{T} >$ 1 GeV/c}; Row~5~--~simulated signals from
PYTHIA-HIJING's mixture of $\pi^{\pm}$, $K^{\pm}$\, and
$p/\overline{p}$\, at \mbox{$P_{T} >$ 1 GeV/c}; Row~6~--~same as
Row~5, but with 0.5~T STAR magnetic field on; Row~7~--~same as
Row~6, but with the background from neutrals equivalent to HIJING's
central Au-Au events.
}
\label{fig:miphst}
\end{figure}

   To evaluate the corrections and the associated systematic errors,
the various hits in the BEMC towers have been simulated, using
GEANT. Then, the energy deposits in the scintillator tiles have been
transformed to the PMT signal distributions, accounting for the photon
and secondary-electron statistics and electronics noise. In the 
simulations, the ``light-yield'' from each single tile varied from 2.5
to 3.5 photoelectrons per perpendicular MIP crossing. Electronic
noise, $\sigma_{noise}$\, varied from 0 to 2.5 ADC channels. PMT gains
have been chosen so as to have MIP peaks\footnote{Mean values.}
sitting in the interval from $\sim$12th to 18th ADC channels above the
pedestal. Typical simulated signal distributions are shown in
Fig.~\ref{fig:miphst}, along with an example of experimental
histograms from the BEMC Beam-Test-'98~\cite{bt98:01}.

It was found that Vavilov distributions~\cite{vavilov:57} quite
satisfactorily approximate GEANT simulated energy deposits from
mono-momentum charged pions with hadronic interactions turned off. To
account also for the photon and secondary-electron statistics in the
PMTs and electronics noise, the following function was used to fit the
respective ADC signal histograms (top left frame of
Fig.~\ref{fig:miphst}):
\begin{equation}
V_{MIP}(x) = a\cdot\int_{PED}^{\infty}
\frac{dy\, \exp\big(-\frac{(x-y)^2}{2\sigma^{2}(y)}\big)}{2\pi\sigma (y)}
\Phi_{V}\Big(\lambda(y);\,\kappa,\beta^{2}\Big)
\label{eq:vavmono}
\end{equation}
where:
\begin{eqnarray}
\Phi_{V}(\lambda;\kappa,\beta^{2}) & & \mbox{is Vavilov distribution
{\em VAVDEN\/} from CERN} \nonumber \\
& & \mbox{library~\cite{vavlov:93};} \nonumber \\
\lambda(y) & = & \frac{y-PED-MIP}{s\cdot
MIP}+\overline{X}_{\Phi}(\kappa,\beta^{2}) ; \nonumber \\
\overline{X}_{\Phi}(\kappa,\beta^{2}) & = &
\frac{\int_{-\infty}^{\infty}\lambda\Phi_{V}(\lambda;\kappa,\beta^{2})\,
d\lambda}
{\int_{-\infty}^{\infty}\Phi_{V}(\lambda;\kappa,\beta^{2})\,d\lambda}\;\;\;
\mbox{is the mean value of a ``bare''} \nonumber \\
& & \mbox{Vavilov distribution}\;
\Phi_{V}(\lambda;\kappa,\beta^{2}); \nonumber \\
\sigma (y) & = & \sqrt{\sigma_{PED}^{2}+\alpha\cdot (y-PED)}\;\;\;
\mbox{is the width of the} \nonumber \\
& & \mbox{smearing gaussian}; \nonumber \\
\mbox{$PED$\, and $\sigma_{PED}$} & & \mbox{are respectively the ADC
pedestal position and} \nonumber \\
& & \mbox{width, known from separate measurements;} \nonumber \\
MIP & & \mbox{is for the MIP-peak position, which is expected}
\nonumber \\
& & \mbox{to be close to the mean value over MIP-peak with}
\nonumber \\
& &\mbox{already subtracted $PED$.} \nonumber
\end{eqnarray}
Constants $\kappa$, $\beta^{2}$\, and $s$\ have been fixed to the
following values:\footnote{The other approach, with $\kappa$,
$\beta^{2}$\, and $s$\, selected individually for each tower's
$\eta$-position and pion momentum from the best $\Phi_{V}$-fits to the
simulated MIP energy deposits, has also been tested. It was found
however that, in the limited momentum range of interest, this approach
does not yield better results compared to the version with fixed
constants.} $\kappa=$~0.3, $\beta^{2}=$~0.98 and $s=$~0.078. Out of 3
parameters, $a$, $MIP$\, and $\alpha$, which were to be determined
from the fits, only one was actually the goal: that was $MIP$.

   Function $\Phi_{V}(\lambda;\kappa,\beta^{2})$\, is quite complex
by itself, and when it is convoluted with a smearing gaussian, the
fitting procedure using $V_{MIP}(x)$\, becomes relatively slow, taking
15--20 seconds per histogram in PAW~\cite{pawguide} with currently
available computers. To make quick $MIP$\, estimates, simple gaussian
fits with 3 parameters: $MIP$, $a$\, and $\sigma_{MIP}$, have also
been evaluated:
\begin{equation}
G_{MIP}(x) =
a\cdot\exp\Big[-\frac{(x-PED-MIP)^2}{2\times\sigma_{MIP}^2}\Big]
\label{eq:gmono}
\end{equation}

   To account for a background from hadronic interactions, etc. under
and in the vicinity of MIP peaks in the signal distributions, a
function with 3 more parameters, $b_{0}$, $b_{1}$\, and $b_{2}$, has
been added to $V_{MIP}(x)$\, and $G_{MIP}(x)$\, in fits to histograms
of Rows~2--7 in Fig.~\ref{fig:miphst}:
\begin{eqnarray}
V(x) & = & V_{MIP}(x) + b_{0}\cdot\exp\big(b_{1}x+b_{2}x^{2}\big)  
\label{eq:vavfull} \\
G(x) & = & G_{MIP}(x) + b_{0}\cdot\exp\big(b_{1}x+b_{2}x^{2}\big)
\label{eq:gfull}
\end{eqnarray}
Fits were usually extended from $PED+3\times\sigma_{PED}$\, to
$PED+2.5\times MIP$.

\begin{figure}[htb]
\centerline{\hbox{
\includegraphics[width=14.3cm,bb=20 330 535 650]{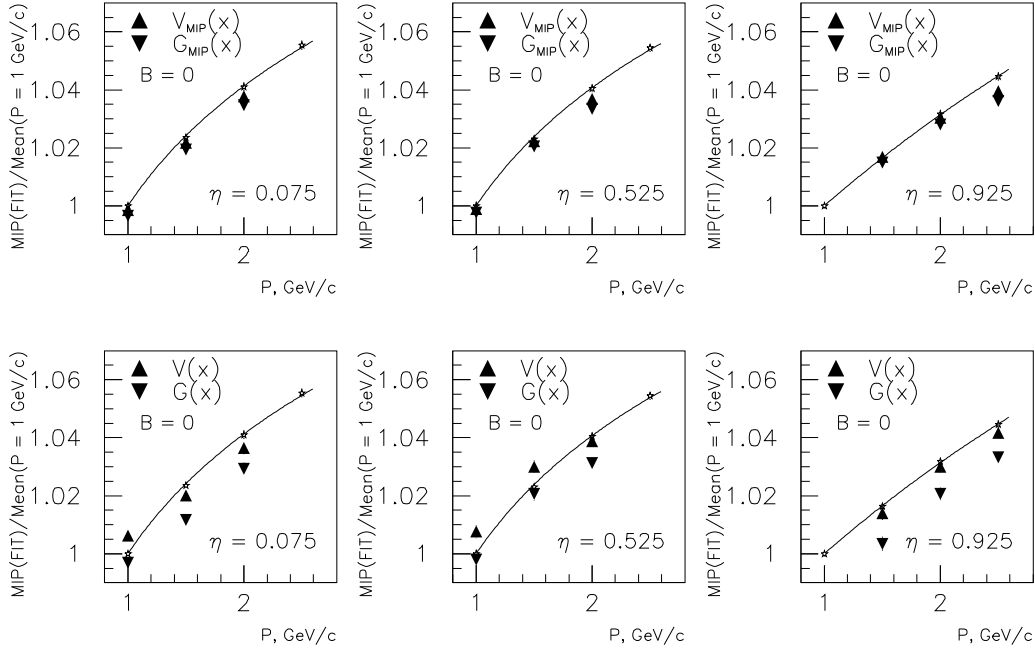}
}}
\caption[]
{
\footnotesize
Comparison of {\em MIP}-parameters from the fits to the means over
signal distributions for pions without hadronic interactions. Upper
row -- 3-parameter fits to histograms from mono-momentum pions without
hadronic interactions (Row~1 of Fig.~\ref{fig:miphst}); Lower row --
6-parameter fits to histograms from mono-momentum pions with hadronic
interactions (Row~2 of Fig.~\ref{fig:miphst}). Momentum dependences
for means are shown with small asterisks and curves.
}
\label{fig:pions}
\end{figure}
In Fig.~\ref{fig:pions}, the results of fits to the signal
distributions from mono-momentum pions are compared to the respective
means over histograms from non-had\-ro\-ni\-cally in\-te\-rac\-ting
pions. Both, $V_{MIP}(x)$\, and $G_{MIP}(x)$\, 3-parameter fits to the 
histograms of Row~1 in Fig.~\ref{fig:miphst} yield $MIP$-values close
to the means and to each other within a fraction of a percent. The
$MIP$s from 6-parameter $V(x)$-fits to the simulated pion signals with
hadronic interactions\footnote{These distributions were expected to be
close to the ones from the BEMC Beam-Test-'98~\cite{bt98:01}, which
are shown in Row~3 of Fig.~\ref{fig:miphst}, because the composition
of the AGS B2 test beam, after electron rejection, using Cherenkov
counters, was mainly $\pi^{-}$-mesons with small admixture of kaons,
antiprotons and muons.} (Row~2 in Fig.~\ref{fig:miphst}) are also in
good agreement with the means, deviating by just about $\pm$0.5\%
(RMS). However in most cases, $G(x)$-fits to the same distributions
underestimate parameter $MIP$\, by approximately 1\%.

   The characteristics of MIP-hit distributions in STAR at RHIC depend
on the hit selection as well as on the characteristics of the selected
events. For track selection in each event, the target towers are those
entered by one and only one charged particle. This particle has
\mbox{$P_{T}\geq P_{T}^{cut}\simeq$~1--1.5~GeV/c}. In the high
multiplicity environment of, for example, central Au-Au events, the
selection conditions could be modified by an additional ``isolation''
requirement of no charged hits in the neighboring towers either. After
checking that an extrapolated track of interest potentially crosses
all 21 scintillator tiles of a tower\footnote{Keeping in mind, of
course, extrapolation uncertainties due to multiple scattering.}, the
tower's signal is added to a respective histogram. All the selections
above are easily made with the STAR tracking system~\cite{star:94}.

\begin{figure}[htb]
\centerline{\hbox{
\includegraphics[width=14.3cm,bb=20 160 535 660]{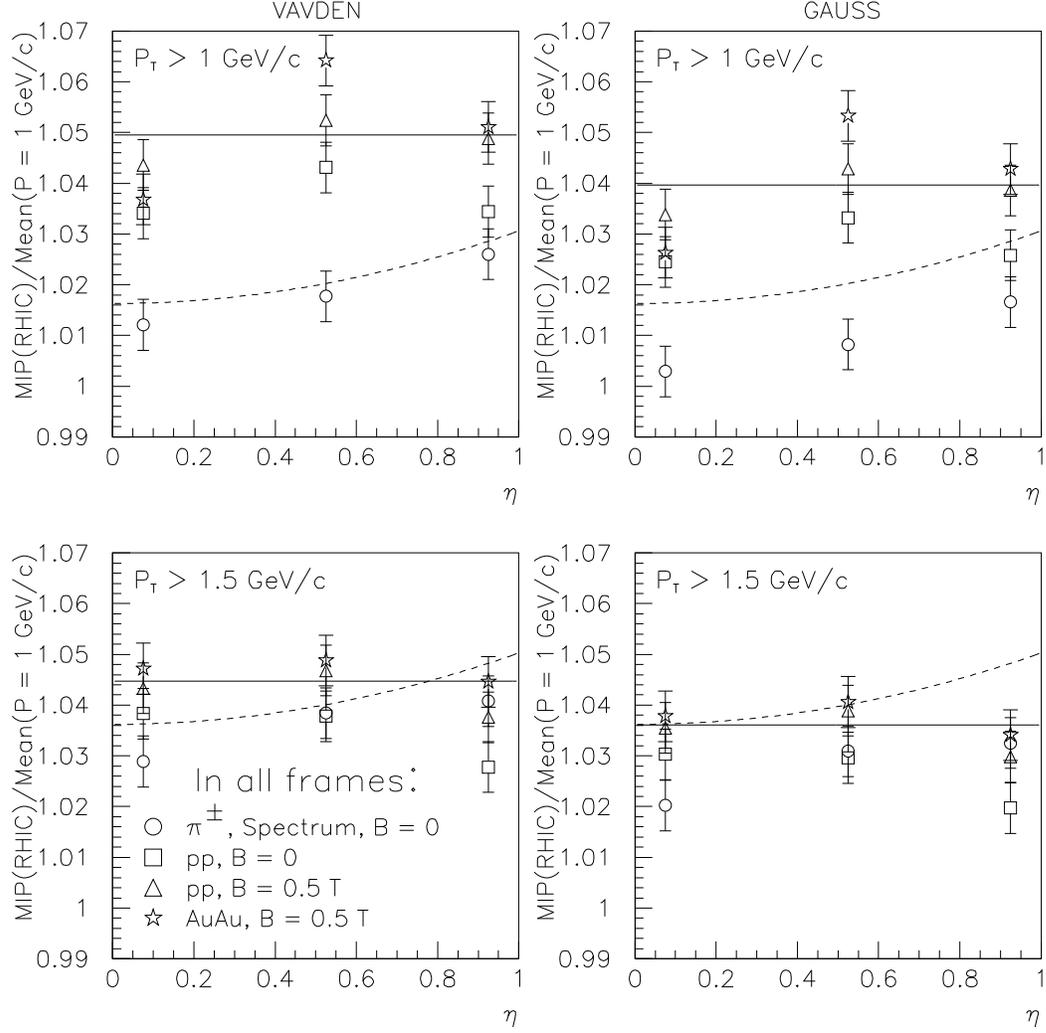}
}}
\caption[]
{
\footnotesize
Results of fits to signal histograms for pions of PYTHIA-HIJING's
$P_{T}$-spectrum ($\pi^{\pm}$, Spectrum, $B =$~0);
PYTHIA-HIJING's mixture of pions, kaons and $p\overline{p}$\,
without ($pp$, $B =$~0) and with ($pp$, $B =$~0.5~T) STAR magnetic
field; HIJING mixture of MIP-hits with the neutral background from
underlying central gold-gold events ($AuAu$, $B =$~0.5~T). See text
for details.
}
\label{fig:rhic}
\end{figure}

   The simulation, which are shown in Rows~4--7 of
Fig.~\ref{fig:miphst}, have been done to evaluate corrections and
systematic errors of MIP-calibrations in STAR at RHIC. The results are
presented in Fig.~\ref{fig:rhic}. At the first step (Row~4 of
Fig.~\ref{fig:miphst}), it has been assumed that all accumulated
MIP-hits were due to pions only, and the contributions to the signals
of invisible hits\footnote{Neutrals and energy leaks from neighbor
towers.} from underlying events have also been neglected. In this case,
the only factor, affecting the positions and shapes of MIP peaks,
is the pion's momentum spread\footnote{Effect of STAR magnetic field
is still ignored here.}. For pion spectra, it would be natural to
expect MIP-peak positions close to that of mono-momentum pions with
$P_{T}$\, equal to the mean over the spectrum,
\mbox{$<P_{T}>=P_{T}^{cut} +$ 0.3 GeV/c}. These expectations are shown
with the dashed lines in Fig.~\ref{fig:rhic}. One can observe that the
results of $V(x)$-fits agree within the errors with the expectations,
while $G(x)$-fits yield $MIP$-parameter values lower by the same
$\sim$1\% as they were in the case of mono-momentum pions with
hadronic interactions on.

   After including the PYTHIA-HIJING's~\cite{pythiaguide} spectra of
kaons and (anti)protons in the MIP samples (Row~5 of
Fig.~\ref{fig:miphst}),\footnote{66\% of pions, 16\% of kaons, and
18\% of $p/\overline{p}$.} $MIP$-parameters move up by approximately
\mbox{1--2\%} for \mbox{$P_{T}^{cut}=$~1 GeV/c}. However, at
\mbox{$P_{T}^{cut}=$ 1.5 GeV/c}, there is virtually no $MIP$\, shifts
beyond the statistical errors. Such a difference between
\mbox{$P_{T}^{cut}=$ 1} and 1.5~GeV/c could be due to some complex
interplay of $MIP(P)$-dependences for all three species which, due to
different masses, reach their minima at different $P$.
Turning on the 0.5~T STAR magnetic field makes the simulated
distributions in Row~6 of Fig.~\ref{fig:miphst} close to the ones,
expected in real minimum bias proton-proton and low-multiplicity
(peripheral) nucleus-nucleus collisions at RHIC. The magnetic field
moves MIP-peak positions up by $\sim$1\% more at
\mbox{$P_{T}^{cut}=$~1 GeV/c}, and by somewhat less values at
\mbox{$P_{T}^{cut}=$~1.5 GeV/c}.

\begin{figure}[htb]
\centerline{\hbox{
\includegraphics[width=14.3cm,bb=20 410 535 650]{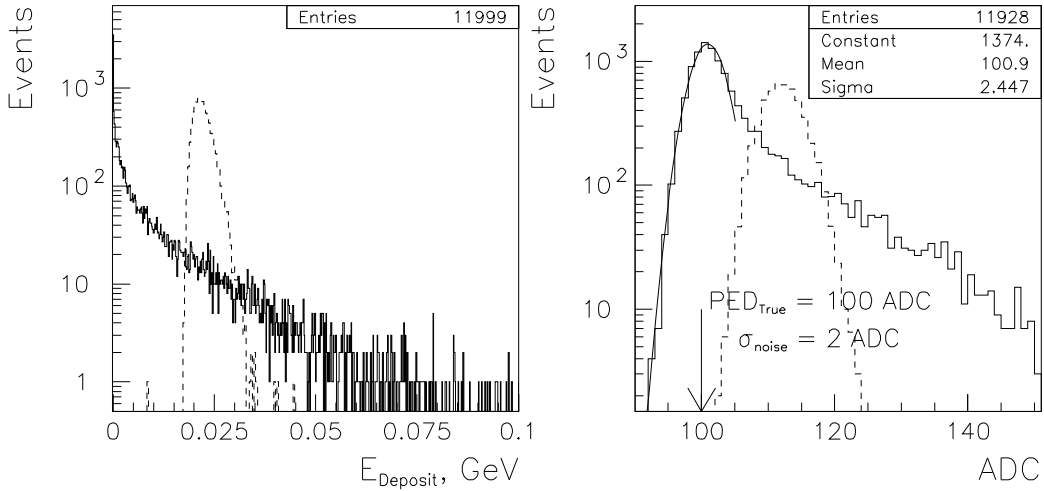}
}}
\caption[]
{
\footnotesize
Typical histogram for HIJING-GEANT energy deposit in a BEMC tower from
neutrals in central Au-Au events (left frame) and the respective PMT
signal distribution in this tower (right frame). The locations of
MIP peaks are marked with the dashed histograms. See text for
details.
}
\label{fig:neutrals}
\end{figure}

To evaluate the effects of invisible hits from underlying events,
the HI\-JING-GEANT energy deposit spectra from neutral particles in
the extreme case of central gold-gold events have been simulated for
the towers at various $\eta$\, (Fig.~\ref{fig:neutrals}, left
frame). Then, by combining these spectra and MIP signal distributions
from Row~6 of Fig.~\ref{fig:miphst}, the histograms for \mbox{Au-Au}
central events (Row~7 of Fig.~\ref{fig:miphst}) have been generated in
the straightforward way, using formula:
\begin{equation}
{\mathcal P}(x) = \int{\mathcal P}_{MIP}(y)\;{\mathcal
P}_{neutral}(x-y)\; dy 
\label{eq:auauprob}
\end{equation}
where ${\mathcal P}(x)$\, is the probability density for the total
signal in a tower, and ${\mathcal P}_{MIP}(y)$\, and ${\mathcal
P}_{neutral}(x-y)$\, are probability densities for the contributions
from MIPs and from invisible neutral hits, respectively.

   It should be underlined that backgrounds from neutrals and, for
example, from hadronic interactions of MIPs themselves affect signal
spectra differently. Hadronic interactions of MIPs add to the spectrum
of purely ionization losses another, much wider distribution, not
shifting in any direction the MIP peak itself. Contrarily, the nonzero
energy deposits from neutrals are added on top of MIP signals, thus
shifting the total signal to the right in virtually every single
event. Actually, large energy deposits from neutrals (larger than MIP
peak width of $\sim$2--3~MeV) are not so dangerous, because they just
reduce the useful statistics, kicking the respective events completely
out of MIP peaks. However, the contributions of small energy deposits
from neutrals are not distinguishable in the
fits~(\ref{eq:vavfull})--(\ref{eq:gfull}) from electronics noise. As a
result, they make the ``effective pedestal'' and MIP peak looking
wider and, due to unipolar nature of this effect, they shift both,
effective $PED$\, and $MIP$\, by about the same amount to the right.

   From the consideration above it follows that, if in
fits~(\ref{eq:vavfull})--(\ref{eq:gfull}), the characteristics of the
effective pedestals were used for $PED$\, and $\sigma_{PED}$\,
rather than of the electronic pedestals measured, for example, with no
beam in RHIC, then these fits should yield better estimates for
$MIP$-parameter even in the high multiplicity environment of central
Au-Au events. These effective pedestals are measurable by accumulating
the signal distributions in each tower from the same sample of events
used for the respective MIP histograms, 
and with the same hit selection criteria except that no charged
particles at all entering a tower. Then, the effective $PED$\, and
$\sigma_{PED}$\, could be extracted from the simple gaussian fits as
shown in the right frame of Fig.~\ref{fig:neutrals}. Using effective
pedestals in the fits to histograms in Row~7 of Fig.~\ref{fig:miphst}
yielded $MIP$-parameters for central Au-Au events in the remarkably
good agreement with $pp$, $B=$~0.5~T (see
Fig.~\ref{fig:rhic})\footnote{Actually, in most cases of measuring the
energy of a single photon or electron, it would be correct to
calculate the net signal relative to an effective pedestal rather than
the electronic one. However in other cases of, for example, measuring
the mean electromagnetic energy in the event, usage of electronic
pedestals would be more appropriate.}.

   The other interesting observation from Fig.~\ref{fig:rhic} is that,
within $\lesssim$1\%, the ratios \mbox{$MIP(RHIC)/Mean(P
=$~1~GeV/c$)$} for the both, $pp$\, and \mbox{Au-Au} events are
virtually independent of either $\eta$\, or $P_{T}^{cut}$, and for the
practical purpose could be fixed at $\sim$1.045--1.050\footnote{For
$V(x)$-fits; using $G(x)$\, suggests the lower values by $\sim$1\%.}
(shown in Fig.~\ref{fig:rhic} with the solid horizontal
lines). Actually, the corrections bring MIP-peak positions in STAR to
approximately \mbox{$MIP(P\simeq$ 2.5 GeV/c$)$} for mono-momentum
pions with no magnetic field. Of course, this combined correction
coefficient is based on simulations only, and, for example, due to
different compositions of MIP-hits in real events, it might require
some additional adjustment. The bottom line, however, is that all
corrections themselves are small, and even if just crudely evaluated
by these simulations, the residual uncertainties should be well below
the limit acceptable for the STAR BEMC calibration errors.

   An important thing to underline is that, in the simulations, all
charged particles originated from the vertex exactly at the STAR's
center. In the real experiment, the vertices are spread over a
significant collision diamond, and as a result, MIP peaks could be
shifted somewhat due to such spread. However, this effect is almost
purely geometrical, and the respective correction can easily be made
to the signal from each hit before filling the histogram.

\begin{figure}[htb]
\parbox{7.4cm}{
\centerline{\hbox{
\includegraphics[width=7.3cm,bb=20 415 280 650]{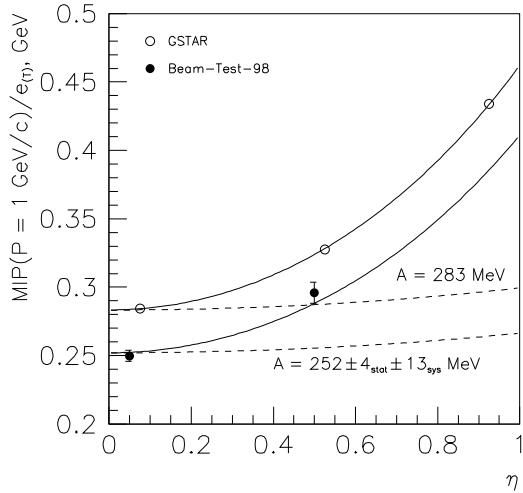}
}}
} \
\parbox{6.0cm}{
\caption[]
{
\footnotesize
$MIP/e_{(T)}$-ratio measured in the BEMC Beam-Test-'98~\cite{bt98:01}
and simulated with GEANT. Solid lines are for $MIP/e$- and dashed
lines are for $MIP/e_{T}$-ratios, where
\mbox{$e_{T}=S_{e}(E_{e})/E_{T}$} and \mbox{$E_{T}=E_{e}\sin\theta$}
(see footnote~\ref{footnote:e} in Sec.~\ref{sec:mipcalib} for other
notations).

}
\label{fig:bt98}
}
\end{figure}
   The \mbox{$MIP(P=$ 1 GeV/c$)/e$-ratios} measured in the
Beam-Test-'98 are shown in
Fig.~\ref{fig:bt98}~\cite{bt98:01}. Functionally, the experimental
$\eta$-dependence of $MIP/e$\, is consistent with GSTAR\footnote{GSTAR
is the configured with STAR setup version of GEANT.} simulations:
\begin{equation}
MIP(P=\mbox{ 1 GeV/c})/e = A\times (1+0.056\eta^{2})/\sin\theta
\label{eq:miptoebt98}
\end{equation}
However, the normalization constants $A$\, differ from each other by
$\sim$12\%. This difference might partially be due to some residual
experimental uncertainties, but overestimating of $MIP/e$\, by the
standard GSTAR at $\sim$5--10\% cannot be excluded either. After
applying the combined  correction coefficient above to the
experimental data, the following formula could be written for the
\mbox{$MIP(RHIC)/e$}\, $\eta$-dependence in STAR:
\begin{equation}
MIP(RHIC)/e = (264\pm 4_{stat}\pm 13_{sys}\mbox{
MeV})\times(1+0.056\eta^{2})/\sin\theta
\label{eq:miptoefin}
\end{equation}
For the time being, until the better data from measurements at an
external beam are available, formula~(\ref{eq:miptoefin}) is suggested
for use in the STAR BEMC calibrations at RHIC.

\section{Statistics and calibration time}
\label{sec:statistics}

   From the analysis of Sec.~\ref{sec:mip_peak} it follows that, using
MIPs from physics events at RHIC allows an absolute calibration of the
BEMC at a percent level of systematic uncertainties. To achieve
comparable statistical errors, a sufficient number of useful MIP-hits
must be accumulated in each tower. MIP-peak widths in signal
distributions are expected at $\sim$20\%. Thus, to measure MIP-peak
positions with a relative statistical error $\delta$, the number of
useful hits should be \mbox{$N_{MIP}\simeq (0.2/\delta)^2$}. Due to
the deflection in the STAR magnetic field, the projected trajectories
of only \mbox{50--60\%} of all charged particles at
\mbox{$P_{T}\simeq$ 1 GeV/c}, which enter a tower, will cross all 21
scintillator tiles. Out of them, \mbox{60--70\%} will experience
hadronic interactions and produce signals out of MIP peaks. As a
results, the total number of \mbox{high-$P_{T}$}\, entries per tower
should be \mbox{$N_{hit}\simeq 6.5\times N_{MIP}\simeq
(0.5/\delta)^2$}.

   At design RHIC luminosity, the limiting factor for the useful event
rate will be the data acquisition bandwidth rather than physics cross
sections. To minimize MIP-hit accumulation time, the STAR Level-3 (L3)
tracking~\cite{lange:00} will be exploited a)~to select among all
incoming events those with \mbox{high-$P_{T}$}\, tracks, pointing out
to the BEMC towers; b)~to reconstruct parameters of these tracks and,
probably, even c)~to fill histograms. The input L3 event rate could be
as high as 100~Hz. However, the actual MIP-hit accumulation rate will
also depend on the type of incoming events. Here we will provide the
time estimates for three event types: a)~$pp$\, and b)~\mbox{Au-Au}
minimum-bias events with at least one charged particle, detected in
the CTB \mbox{($N_{CTB}\geq 1$)} within its acceptance of
$-1\leq\eta\leq 1$, $0\leq\varphi\leq 2\pi$, and c)~low-multiplicity
(peripheral) \mbox{Au-Au} events with $1\leq N_{CTB} \leq 100$. In
these events, according to PYTHIA-HIJING, the average numbers of
tracks with \mbox{$P_{T} >$ 1 GeV/c}\, within the BEMC acceptance
would be 0.23, 7.25 and 1.05, respectively. Putting these numbers
together, we get the running times of Table~\ref{tab:runtime}, to
achieve the indicated statistical accuracies on the MIP-peak
positions.
\begin{table}[htb]
\caption{\centerline{RHIC running time and statistical errors for
MIP-calibrations}}
\label{tab:runtime}
\begin{center}
\begin{tabular}{||l||c|c|c|c|c||} \hline\hline
Statistical error & 20\% & 10\% & 5\% & 2\% & 1\% \\ \hline\hline
Minimum Bias $pp$ & 23 min & 1.5 hr & 6 hr & 38 hr & 6.3 day \\
Peripheral Au-Au & 5 min & 20 min & 1.3 hr & 8.3 hr & 1.4 day \\
Minimum Bias Au-Au & 1 min & 3 min & 12 min & 1.2 hr & 5 hr \\
\hline\hline
\end{tabular}
\end{center}
\end{table}

   It is worth noting that the above estimates are for the time
required to get every BEMC tower calibrated. However, after the towers
are equalized to a few percents, the monitoring of the mean gain
variation for a patch, consisting of $N_{tw}$\, towers, will take by a
factor of $N_{tw}$\, less time compared to what was necessary to
calibrate every single tower with the same statistical uncertainty.

\section*{Conclusion}
\label{sec:conclusion}

   To summarize, the study in this paper has shown that, using
MIP-hits, the equalization and transfer of the absolute scale from the
test beam calibrations can be done to a percent level accuracy in a
reasonable amount of time for the entire STAR BEMC. MIP-hits are
also an effective tool for continuously monitoring the variations of
the BEMC tower's gains, virtually without interference to STAR's main
physics program. This method does not rely on simulations for anything
other than geometric and some other small corrections, and also for
estimations of the systematic errors: it directly transfers measured
test beam responses to operations at RHIC.

\section*{Acknowledgments}

   It is our pleasure to thank E.~P.~Kistenev, J.~M.~Landgraf,
J.~S.~Lange, M.~J.~LeVine, P.~L.~Nev\-ski, E.~K.~Shabalina,
P.~P.~Yepes and many members of the STAR EMC collaboration for the
useful discussions. This work has been supported in part by the
U.S. Department of Energy Grant DE--FG0292ER40713.

\end{document}